\def \SAIT #1 #2 {{\em Mem.\ Soc.\ Astron.\ It.\/} {\bf #1}, #2}
\def \MESS #1 #2 {{\em The Messenger\/} {\bf #1}, #2}
\def \ASTRNACH #1 #2 {{\em Astron. Nach.\/} {\bf #1}, #2}
\def \AAP #1 #2 {{\em Astron. Astrophys.\/} {\bf #1}, #2}
\def \AAL #1 #2 {{\em Astron. Astrophys. Lett.\/} {\bf #1}, L#2}
\def \AAR #1 #2 {{\em Astron. Astrophys. Rev.\/} {\bf #1}, #2}
\def \AAS #1 #2 {{\em Astron. Astrophys. Suppl. Ser.\/} {\bf #1}, #2}
\def \AJ #1 #2 {{\em Astron. J.\/} {\bf #1}, #2}
\def \ANNREV #1 #2 {{\em Ann. Rev. Astron. Astrophys.\/} {\bf #1}, #2}
\def \APJ #1 #2 {{\em Astrophys. J.\/} {\bf #1}, #2}
\def \APJL #1 #2 {{\em Astrophys. J. Lett.\/} {\bf #1}, L#2}
\def \APJS #1 #2 {{\em Astrophys. J. Suppl.\/} {\bf #1}, #2}
\def \APSS #1 #2 {{\em Astrophys. Space Sci.\/} {\bf #1}, #2}
\def \ASR #1 #2 {{\em Adv. Space Res.\/} {\bf #1}, #2}
\def \BAIC #1 #2 {{\em Bull. Astron. Inst. Czechosl.\/} {\bf #1}, #2}
\def \JSQRT #1 #2 {{\em J. Quant. Spectrosc. Radiat. Transfer\/} {\bf #1}, #2}
\def \JAA #1 #2 {{\em J. Astrophys. Astr.\/} {\bf #1}, #2}
\def \MN #1 #2 {{\em Mon. Not. R. Astr. Soc.\/} {\bf #1}, #2}
\def \MEM #1 #2 {{\em Mem. R. Astr. Soc.\/} {\bf #1}, #2}
\def \PLR #1 #2 {{\em Phys. Lett. Rev.\/} {\bf #1}, #2}
\def \PASJ #1 #2 {{\em Publ. Astron. Soc. Japan\/} {\bf #1}, #2}
\def \PASP #1 #2 {{\em Publ. Astr. Soc. Pacific\/} {\bf #1}, #2}
\def \NAT #1 #2 {{\em Nature\/} {\bf #1}, #2}

\documentstyle{memsait}
\input epsf.sty
\begin{opening}
\title{POLARIZATION AND THE AGE OF PULSARS} 
\author{Alexis von Hoensbroech}
\institute{Max-Planck-Institut f\"ur Radioastronomie,
 Auf dem H\"ugel 69, D-53121 Bonn, Germany.}
\date{} 
\end{opening}

\begin{document}

\oddpagefooter{}{}{} 
\evenpagefooter{}{}{} 
\ 
\bigskip

\begin{abstract}
Using polarimetric radio profiles of 84 pulsars at a wavelength 
of 6 cm, we present a correlation between the polarization and
characteristic age of pulsars.
Considering the large error connected with the value of the 
characteristic age, we speculate about the possibility
of identifying young pulsars through their polarimetric properties.
Using this method we have identified more than 20 possibly young
pulsars including some with a relatively high characteristic age. 
Of these objects, seven have already been proposed to be associated
with supernova remnants. For the remaining pulsars we find a number of
candidates for associated faint SNRs and present one of them. This
shows that this method can be used as an additional independent
indicator for the youthfulness of pulsars. 
\end{abstract}

\section{The Pulsar Age Problem}

Although it is commonly believed that pulsars are born in the supernova
explosions of massive stars, only about 2\% of the presently observed
population are considered to be associated with supernova remnants (hereafter
SNRs). The standard explanation for this apparently low fraction of
objects is a combination of the radio lifetimes of pulsars ($\sim 10^7$ yr) 
compared to SNRs ($\sim 10^5$ yr), the beaming fraction of radio
pulsars and the 
fraction of all supernova that produce a neutron star. In order to decide
whether an association is real or coincidental, a number of questions need
to be considered (see e.g. Kaspi 1996). These include positional
coincidence, agreement of the distances and ages of both the pulsar
and the SNR
and, if measured, the velocity vector of the pulsar should be consistent
with its position offset from the center of the SNR. Agreement in age is
of great importance, since the age of the pulsar should not much exceed
$10^5$ years. 

The age of a pulsar can in principle be determined from its spin
parameters. Given an initial spin period $P_{\rm i}$ and a magnetic
field $B$, the pulsar loses rotational energy,
mainly through magnetic dipole radiation, but also through
electromagnetic radiation and particle losses. The spin period $P$ 
therefore increases with time. As the temporal derivative of the
period $\dot P$ 
can be measured with a high accuracy, the age of the pulsar can be
determined if assumptions are made for $P_{\rm i}$ and the braking
behavior (e.g. Manchester \& Taylor 1977).
The equation of motion is described by the differential equation
\begin{equation}
\dot\Omega=-K\Omega^{n},
\end{equation}
where $\Omega=2\pi/P$ and $K$ is a constant. The braking index $n$ 
depends on the specific slow-down model. Integrating
this equation yields an expression for the age:
\begin{equation}
T=\frac{1}{n-1}\frac{P}{\dot P}\left[1-\left(\frac{P_{\rm
i}}{P}\right)^{n-1}\right].
\end{equation}
It is a standard practice to define the 
``{\it characteristic age}'' $\tau$, where $P_{\rm i}$ is assumed to
be 0 (corresponding to $P_{\rm i} \ll P$) and $n=3$ (corresponding to
the braking torque through pure dipole radiation)
\begin{equation}
\tau=\frac{P}{2\dot P}\ .
\end{equation}

\begin{figure}
\epsfxsize=13cm
\hspace{1.4cm}\epsfbox{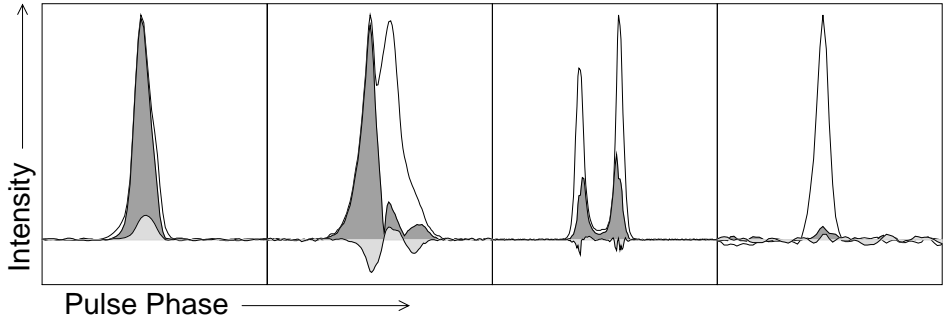}
\caption[h]{Polarization profiles of four different pulsars. The
enveloping solid line is the total power, the dark and the 
light shaded areas correspond to the linear and circular polarized
power respectively. Basically all polarization states can be
found. Examples are given for fully-, partly-, semi- and unpolarized
profiles.} 
\end{figure}

In principle, $n$ can be calculated through
measurement of $\ddot P$. In practice this is difficult due to the
presence of timing noise and glitch activity for young pulsars.
Nevertheless $\ddot P$ has been determined for four pulsars so 
far and varies between n=1.4 for the Vela pulsar and 2.8 for PSR
B1509-58 (Manchester et al. 1985, Lyne et al. 1988, Manchester \&
Peterson 1989, Lyne et al. 1996). As can be seen in
Eq. (2) this span already implies an uncertainty of a
factor of about 4.5 for $T$. All values are smaller than $n=3$ which
leads to a
systematic underestimation by the characteristic age. This
underestimation is important for long-period pulsars with $P\gg P_{\rm i}$.

Estimates for $P_{\rm i}$ on the other hand are mainly
based on theoretical models. An exception is the Crab pulsar where the
age is precisely 
known through Chinese records. The corresponding supernova explosion
occurred in 1054 A.D.. Using Eq. (2), the initial period of this
pulsar can
therefore be calculated as $P_{\rm i}=19.3 {\rm ms}$. In the
literature, estimates for
the initial period of pulsars are
found between virtually zero and about $P_{\rm i}=500 {\rm ms}$
(e.g. Lyne et al. 1985, Narayan \& Ostriker 1990 and Lorimer et
al. 1993). Therefore for
short period pulsars, $P_{\rm i}$ could well be of the order of
the present period. 
Therefore the characteristic age could
easily {\it overestimate} the real age of these pulsars by more than an
order of magnitude (Kaspi et al. 1997).

\section{Relation to Pulsar Polarization}

\begin{figure}[t]
\epsfxsize=12cm
\hspace{0cm}\epsfbox{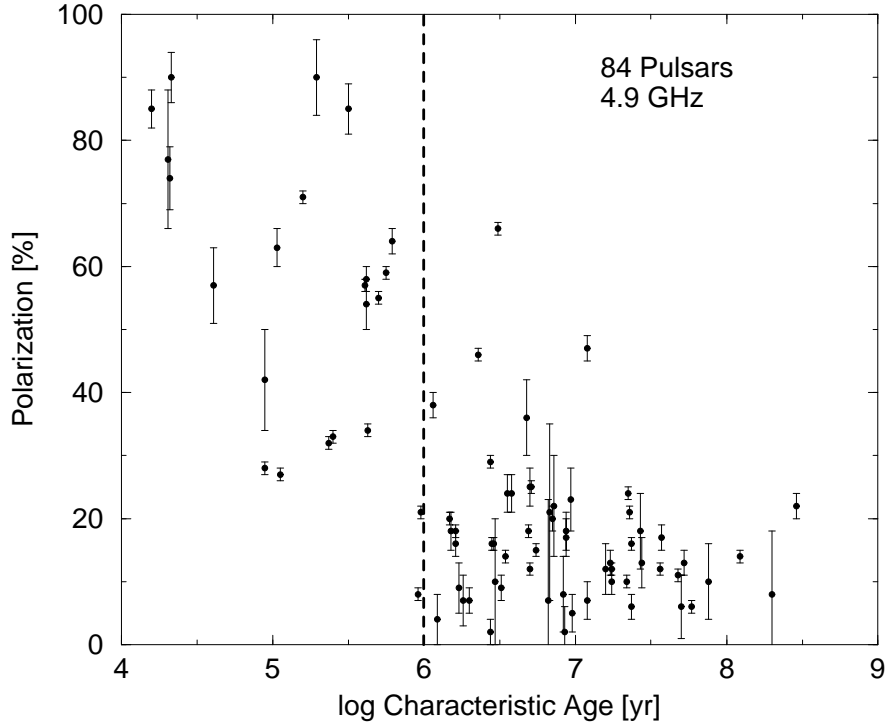}
\caption[h]{Degree of polarization $\Pi$ versus the characteristic age
$\tau$. Young pulsars with $\tau < 10^6$ yrs are significantly higher
polarized than old objects. The possibly large error connected with
the characteristic age induces a strong scatter in the horizontal
direction.}
\end{figure}

The examples of pulsar polarimetry in Fig. 1 show that basically all states
of polarization can be found.
One question is, if the degree of
polarization (hereafter $\Pi$), which is defined as the polarized
fraction of the total intensity, depends on any pulsar parameter at all. 
Although it was 
known that young pulsars often have highly polarized profiles
(e.g. Manchester 1996), no significant correlations could be found so far. As
demonstrated by von Hoensbroech et al. (1998a) this is mainly due to
the fact that earlier investigations had been made at ``classical''
pulsar frequencies between 400 MHz and 1400 MHz. 
At frequencies higher than a few GHz clear correlations become
obvious between $\Pi$ and the loss of rotational
energy $\dot E$, respectively the surface accelerating potential. 

At 4.9 GHz, for instance, pulsars with a high
$\dot E$ clearly have a higher $\Pi$ than pulsars
with a low $\dot E$ (for a possible physical explanation see von
Hoensbroech et al. 1998b). As $\dot E$ decreases with the
characteristic age, $\Pi$ also decreases (see
Fig. 2). The potentially large error of the characteristic age 
induces a strong scatter in the horizontal direction of Fig. 2,
making the correlation less significant than it is with $\dot E$. But
one clear conclusion can be drawn from this plot: Young pulsars are
higher polarized than old ones at 4.9 GHz.
Using this fact, we use the polarization of the 4.9 GHz radio profiles
to identify sources with a relatively high characteristic age, which
are possibly younger than they seem.

\section{Results}

\begin{figure}[h]
\epsfxsize=8cm
\hspace{2cm}\epsfbox{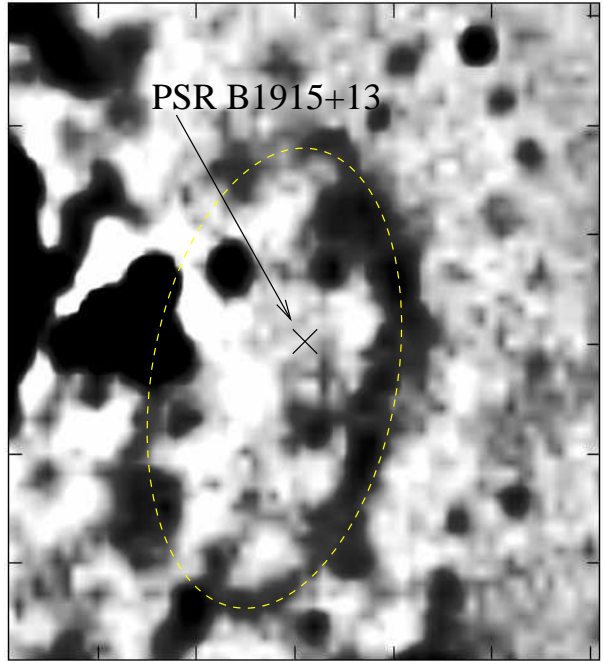}
\caption[h]{Faint shell structure around PSR B1915+13 at $\lambda=11$ cm. 
$2^\circ\times2^\circ$--map from Reich et al. (1990). Accessible
through {\it http://www.mpifr-bonn.mpg.de/survey.html}.}
\end{figure}

\vspace{1cm} 
\begin{table}[h]
\centerline{\bf Tab. 1 - Highly polarized pulsars at 4.9 GHz}
\hspace{0.5cm} 
\begin{tabular}{|l|c|c|c|c|l|}
\hline
Pulsar&l/b[$^{\circ}$]&P[ms]&$\tau$ [log yr]&$\Pi$ [\%]&Remarks \& Ref.\\
\hline
\hline
B0136+57 & 129.2/$-$4.0 & 272 & 5.6 & $57\pm 1$&\\
B0144+59 & 130.1/$-$2.7 & 196 & 7.1 & $47\pm 2$&\\
B0355+54 & 148.2/0.8 & 156 & 5.8 & $59\pm 1$&\\
B0450+55 &152.6/7.5 & 341 & 6.4 & $46\pm 1$&\\
J0538+2817 & 179.9/$-$1.7 & 143 & 5.8 & $63\pm 2$&association [1]\\
\hline
B0540+23 & 184.4/$-$3.3 & 246 & 5.4 & $33\pm 1$&\\
B0559$-$05 & 212.2/$-$13.5 & 396 & 6.7 & $36\pm 6$&no map\\
B0611+22 & 188.8/2.4 & 335 & 5.0 & $42\pm 8$&association [2]\\
B0740$-$28 & 243.8/$-$2.4 & 167 & 5.2 & $71\pm 1$&no map\\
B0833$-$45 & 263.6/$-$2.8 &  89 & 4.1 & $70\pm 5$&association [3]\\
\hline
B0919+06 & 225.4/36.4 & 431 & 5.7 & $55\pm 1$&\\
B1702$-$19 & 3.2/13.0 & 299 & 6.1 & $38\pm 2$&\\
B1737$-$30 & 358.3/0.2 & 607 & 4.3 & $74\pm 5$&strong background\\
B1800$-$21 & 8.4/0.1 & 134 & 4.2 & $85\pm 3$&association [4]\\
B1822$-$14 & 16.8/$-$1.0 & 279 & 5.4 & $90\pm 6$&association [5]\\
\hline
B1822$-$09 & 21.4/1.3 & 769 & 5.4 & $32\pm 1$&\\
B1823$-$13 & 18.0/$-$0.7 & 102 & 4.3 & $91\pm 4$&strong background\\
B1828$-$10 & 20.8/$-$0.5 & 405 & 5.0 & $63\pm 3$&strong background\\
B1853+01 & 34.6/$-$0.5 & 267 & 4.3 & $77\pm 11$&association [6]\\
B1913+10 & 44.7/$-$0.6 & 405 & 5.6 & $54\pm 4$&\\
\hline
B1915+13 & 48.3/0.6 & 195 & 5.6 & $34\pm 1$&SNR candidate\\
B1929+10 & 47.4/$-$3.9 & 227 & 6.5 & $66\pm 1$&\\
B2011+38 & 75.9/2.5 & 230 & 5.6 & $58\pm 2$&strong background\\
B2334+61 & 114.3/0.2 & 495 & 4.6 & $57\pm 6$&association [7]\\
\hline
\end{tabular}
\caption[h]{Highly polarized pulsars at 4.9 GHz. All data were taken with
the Effelsberg 100m radio telescope except for B0833-45 (Komesaroff et
al. 1974). Known associations with SNRs are marked. It is not ruled
out that some of them are chance associations. References are: [1]
Anderson et al. (1996), [2] Davies et al. (1972), [3] Large et
al. (1968), Kassim \& Weiler (1990), [5] Reich et al. (1986), [6]
Wolszczan et al. (1991), [7] Kulkarni et al. (1993).}
\end{table}

Among the 84 pulsars for which 4.9 GHz data were available (Komesaroff et
al. 1974, von Hoensbroech \& Xilouris 1997, von Hoensbroech et al.
1998a), 24 have a degree of polarization exceeding 30\%. 
All pulsars and their parameters are listed in Tab. 1.
For two pulsars we could not 
examine high resolution radio maps of their surroundings, another four
lie in regions of strong Galactic background emission which makes it difficult
to detect possible faint SNRs. Among the remaining 18 pulsars we find 7
known proposed associations with SNRs -- {\it corresponding to a
fraction of $\sim$40\%} 
-- and a couple of SNR-candidates. This is far above the fraction of
pulsar-SNR associations in the whole pulsar sample which does not
exceed $\sim$2\%.

Fig. 3 shows a faint shell structure which we found around PSR B1915+13
at a wavelength of $\lambda=11$-cm. The data were taken from the
Effelsberg 11cm Galactic plane survey (Reich et al. 1990). 
Large scale diffuse background emission has been subtracted appropriately.
This pulsar has a spin period of $P=195$ ms and a characteristic age of 
$\tau=4\cdot10^5$ yrs. If this would be the pulsars true age an 
associated SNR should have vanished. The pulsar radiation is about 34\% 
polarized at 4.9 GHz which could indicate a smaller true age.
It is not yet clear if the shell structure around this pulsar 
is of non-thermal origin, but a verifying observation will be carried 
out soon.

The results show that the selection of highly polarized pulsars at
about five GHz is a powerful method to identify possibly young
pulsars. We stress that the method is totally independent of any other
pulsar parameter and 
could be particularly useful to choose additional candidates for the
search of associated faint SNRs.


\acknowledgements
I want to thank W. Reich for invaluable help with the radio maps
and interesting discussions. I also thank R. Wielebinski and
D. Lorimer for their support, interest and comments on the manuscript.

\end{document}